\documentclass[%
 reprint, 
 superscriptaddress,
 amsmath,amssymb,
 aps,
floatfix,
longbibliography
]{revtex4-2}
\usepackage{graphicx} 
\usepackage{amsmath}
\usepackage[colorlinks=true,citecolor=blue,linkcolor=magenta]{hyperref}
\usepackage{braket}
\usepackage{enumitem}

\allowdisplaybreaks 
\usepackage{float}
\begin{document}

\title{Reducing Quantum Error Mitigation Bias Using Verifiable Benchmark Circuits}
\hypersetup{pdftitle={Reducing Quantum Error Mitigation Bias Using Verifiable Benchmark Circuits}}
\author{Joseph Harris}
\email{joseph.harris@dlr.de}
\author{Kevin Lively}
\author{Peter K.~Schuhmacher}

\affiliation{Department of High Performance Computing, Institute of Software Technology, German Aerospace Center (DLR),  Rathausallee 12, 53757 Sankt Augustin, Germany}
\date{March 2026}

\begin{abstract}

We present a simple, malleable and low-overhead approach for improving generic biased quantum error mitigation (QEM) methods, achieving up to 15\% fidelity improvements over standard QEM on 100-qubit circuits with up to 2000 entangling gates. We do so by constructing verifiable \textit{benchmark circuits} which mirror the application circuit's native-gate structure and thus noise profile. These circuits can be used to benchmark and mitigate the bias of the underlying error mitigation method, requiring only the application circuit and hardware native gate set. We present two methods for generating benchmark circuits; one is agnostic to the target hardware at the expense of a small overhead of single-qubit gates, while the other is specific to the IBM superconducting hardware and has no gate overhead. As a corollary, we introduce \textit{benchmarked-noise zero-noise extrapolation} (bnZNE) as a simple adaptation of zero-noise extrapolation (ZNE), one of the most popular error mitigation methods. We consider as an example the bias-mitigated ZNE and bnZNE of Trotterized Hamiltonian simulations, observing that our approaches outperform standard ZNE using both small-scale classical simulations and 100-qubit utility-scale experiments on the IBM superconducting hardware. We consider the measurement of both single-site observables as well as two-site correlations along a one-dimensional qubit chain. We also provide a software package for implementing the error mitigation techniques used in this research.
\end{abstract}

\maketitle

\section{Introduction}
Quantum error mitigation (QEM) \cite{Strikis2021,Suzuki2022,Endo2018,Endo2021,Takagi2022, filippov2023scalabletensornetworkerrormitigation, RevModPhys.95.045005} refers to the class of techniques used to mitigate errors on noisy intermediate-scale quantum (NISQ) computers \cite{Preskill2018quantumcomputingin} where there are too few qubits available and gate fidelity is too poor to allow the use of quantum error correction. Rather than active measurement and feedback, QEM runs the same logical circuit, or a class of modified circuits, a large number of times and uses classical post-processing to improve the accuracy of expectation values. These methods typically suffer from various limitations to their scalability, such as a trade-off between bias and variance \cite{Endo2021}, an overhead of additional circuit runs which ultimately scales exponentially with the circuit size \cite{Takagi2022}, or extremely sophisticated and costly noise channel characterization and post-processing on high-performance computers \cite{Fischer2026}. 

\begin{figure}[]
    \includegraphics[scale=1]{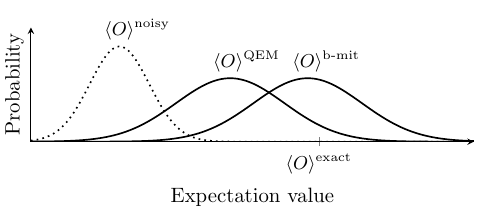}
    \caption{
        Many error mitigation methods result in a less-biased expectation value distribution with greater variance compared to the unmitigated noisy distribution. Our work is to benchmark and mitigate the remaining bias, producing a bias-mitigated (`b-mit') distribution.
     }
    \label{fig:bias-variance}
\end{figure}

In this work, we present methods to improve the accuracy of this first class of biased error mitigation schemes.
Given an input application circuit, we define a corresponding family of \textit{benchmark circuits} which mimic the hardware-level gate structure and thus error behaviour of the application circuit, while still being classically verifiable (e.g.\ Clifford). We show that these circuits can be used to benchmark and then mitigate the bias of an error mitigation method to produce a \textit{bias-mitigated} distribution, as depicted in Figure \ref{fig:bias-variance}.

Given such benchmark circuits, we show analytically that when compensating for the effect of uniform depolarizing noise, our method improves upon the bias of standard digital zero-noise extrapolation (ZNE) \cite{Giurgica_Tiron_2020, 10313813, PhysRevX.7.021050, PhysRevLett.119.180509, Kandala2019, PhysRevA.102.012426}, one of the most popular error mitigation methods due to its relative low-cost and ease of implementation. On top of this, we define an improved ZNE method, \textit{benchmarked-noise zero-noise extrapolation (bnZNE)}, inspired by recent adaptations of ZNE which aim to improve the classical extrapolation process by benchmarking error rates \cite{PhysRevA.110.042625, kim2025enhancedextrapolationbasedquantumerror}. Finally, we show both numerically and experimentally that our methods give more accurate results by applying bias-mitigated ZNE and bnZNE to the Hamiltonian simulation of the kicked Ising and Heisenberg models. We run both small-scale experiments with classical noisy simulation as well as large-scale 100-qubit experiments on the \verb{ibm_fez{ superconducting device, in both cases demonstrating that our methods produce more accurate results compared to regular ZNE. We consider both single-site $\braket{Z}$ expectation values as well as measuring the exponential decay of two-qubit correlations with their distance along a one-dimensional Ising chain.

We present two methods of generating benchmark circuits. The first is intentionally hardware-agnostic and considers the application circuit in terms of Pauli rotations. By assuming that any given Pauli rotation may be transpiled to hardware with comparable errors, it is possible to generate circuits which mimic the application circuit structure but have expectation values verifiably equal to one, typically at the cost of a small overhead of single-qubit gates when compared to the most efficient transpilation of the application. For the second method, we show how more efficient benchmark circuits can be constructed using knowledge of the target hardware, considering the specific example of the IBM `Heron' architecture of superconducting hardware with basis gate set $\{ \text{CZ}, R_Z(\theta), X, \sqrt{X}\}$. In this case, the gate overhead can be reduced to zero. Our approach is intentionally modular -- we present two methods for generating benchmark circuits, but it is possible to derive other methods depending on the specific target application and hardware. We discuss this at the end of the paper and in the appendix. 

The use of classically verifiable structure-mirroring circuits to improve error mitigation through noise estimation has been explored before. One previous work considers zero-noise extrapolation applied specifically to CNOT circuits \cite{PhysRevLett.127.270502}. Our work, however, can be applied to any generic biased error mitigation procedure for hardware with an arbitrary basis gate set under simple assumptions. Our previous work also uses benchmark circuits to detect fluctuations in time-dependent noise in the measurement-based model \cite{PhysRevA.111.022602}.

This work is structured as follows. In section \ref{sec:bias-mitigation-using-benchmark-circuits} we first introduce some simple constraints that benchmark circuits must satisfy and show how they can be used to mitigate the bias of an error mitigation method, given as input an application circuit and target hardware platform or basis gate set. We analytically prove a reduction in bias for the case of ZNE under uniform depolarizing noise channels. We present our benchmarked-noise adaptation of ZNE, bnZNE, in section \ref{sec:benchmarked-noise-zne}.

We then provide two methods for generating benchmark circuits in section \ref{sec:generating-benchmark-circuits}; the first is hardware-agnostic and the second hardware-tailored. In section \ref{sec:numerical-results}, we demonstrate our methods both numerically and on quantum hardware by considering two applications. Finally, we conclude and discuss the outlook of our work in section \ref{sec:discussion}. We also include further details and discussions in the appendix.

Alongside this paper, we also provide a software package for implementing the techniques used in this work as well as containing the numerical data \cite{Github}.


\section{Bias mitigation using benchmark circuits.} 
\label{sec:bias-mitigation-using-benchmark-circuits}
The ultimate objective throughout this work is to produce an estimate to the expectation value $\braket{O}_\mathcal{A}^\text{exact}$ of an observable $O$ with respect to an $n$-qubit \textit{application circuit} $\mathcal{A}$, given access to some quantum error mitigation method and hardware platform. Without loss of generality, a generic observable $O$ can be written as a linear combination of Pauli strings $P$,
\begin{equation}
    O \ = \sum_{P \in \{I,X,Y,Z\}^{\otimes n}} \alpha_P P
\end{equation}
with each $\alpha_P \in \mathbb{C}$. By the linearity of expectation values, it suffices to measure the individual Pauli expectation values separately:
\begin{equation}
    \braket{O}_\mathcal{A} \ = \sum_{P \in \{I,X,Y,Z\}^{\otimes n}} \alpha_P \braket{P}_\mathcal{A}.
\end{equation}
Hence, for simplicity, we adopt the usual assumption that our input observable is a single Pauli string $O\in\{I,X,Y,Z\}^n$. We also assume that all circuits start with the all-zero input state and do not contain mid-circuit measurements.

Given an application circuit $\mathcal{A}$, we define its corresponding benchmark circuit $\mathcal{B} \equiv \mathcal{B}(\mathcal{A})$. The benchmark circuit may be fully determined or part of a family of circuits from which we sample randomly. To keep things as general as possible, we state that $\mathcal{B}$ must satisfy the following conditions:
\begin{itemize}
    \item[C1:] Circuits $\mathcal{A}$ and $\mathcal{B}$ may be transpiled to the target hardware such that they contain the same sequence of native gates and measurements, up to replacing existing gates with new equivalent-error gates.
    \item[C2:] The measurement outcome of $\mathcal{B}$ is a single bitstring which can be efficiently classically determined.
    \item[C3:] The expectation value of $O$ under $\mathcal{B}$ is unity: $\braket{O}_\mathcal{B}^\text{exact} = 1$.
\end{itemize}

In reality, C3 may be satisfied by benchmark circuits with $\braket{O}_\mathcal{B}^\text{exact} = -1$, provided we perform a classical bit-flip on the measurement outcomes to flip the sign of the expectation value. Indeed, the condition $\braket{O}_\mathcal{B}^\text{exact} = \pm 1$ holds for any benchmark circuit satisfying C2. 

As mentioned in the introduction, our definition of a benchmark circuit is intentionally broad so that our method remains modular. In other words, it is possible to come up with different benchmark circuits which may be tailored to particular applications, hardware platforms or error mitigation methods. In section \ref{sec:generating-benchmark-circuits} we give two broad methods of generating benchmark circuits, and we give a further tailored method for the simulation of Trotterized Hamiltonians in the appendix.  First, we show how these circuits can be used for bias mitigation and state our adapted bnZNE protocol.

\subsection{Bias mitigation of arbitrary error mitigation methods}
\label{bias-mit-subsec}


Suppose we are given a general error mitigation scheme which (under some given noise assumptions) corrects errors up to order $\mathcal{O}(p^m)$ ($m\geq 1$) under some (e.g.\ depolarizing) noise model parameterized by error rate $p \ll 1$. We can then write the error-mitigated \textit{expectation value fidelity} of our application circuit as
\begin{equation}
    \frac{\braket{O}_\mathcal{A}^\text{QEM}}{\braket{O}_\mathcal{A}^\text{exact}} \equiv 1 + b_\mathcal{A} = 1 + \mathcal{O}(p^m)
    \label{eq:exp-val-fid}
\end{equation}
where $\braket{O}_\mathcal{A}^\text{QEM}$ is the estimate of $\braket{O}_\mathcal{A}^\text{exact}$ obtained by running our error mitigation method on $\mathcal{A}$, and we call $b_\mathcal{A}$ of order $\mathcal{O}(p^m)$ the \textit{fidelity bias}. By construction, an analogous expression holds for the (averaged) fidelity of the benchmark circuits,
\begin{equation}
    \braket{O}_\mathcal{B}^\text{QEM} \equiv 1 + b_\mathcal{B} = 1 + \mathcal{O}(p^m)
\end{equation}
where we drop the denominator on the left-hand side since $\braket{O}_\mathcal{B}^\text{exact}=1$ for the benchmark circuits by condition C3 above. Then, we can attempt to reduce the fidelity bias of the error-mitigated application circuit's expectation value by computing the \textit{bias-mitigated} quantity,
\begin{align}
    \braket{O}_\mathcal{A}^\text{b-mit} &\equiv  \frac{\braket{O}_\mathcal{A}^\text{QEM}}{\braket{O}_\mathcal{B}^\text{QEM}} 
    \label{eq:bias-mit}\\ &= \braket{O}_\mathcal{A}^\text{exact}\frac{1+b_\mathcal{A}}{1+b_\mathcal{B}} \\
    &= \braket{O}_\mathcal{A}^\text{exact}\left( 1+(b_\mathcal{A} - b_\mathcal{B}) + \mathcal{O}(p^{2m}) \right).
    \label{eq:b-mit}
\end{align}
The last line uses the Taylor series expansion of $1/(1+b_\mathcal{B})$ and multiplies out the terms. The  quantity in equation (\ref{eq:b-mit}) has fidelity bias 
\begin{equation}
     b_\mathcal{A} - b_\mathcal{B} + \mathcal{O}(p^{2m}).
\end{equation}


The assumption we hence need to justify is that
 \begin{equation}
     |b_\mathcal{A} - b_\mathcal{B}| \leq |b_\mathcal{A}|
 \end{equation}
 such that our bias-mitigated expectation value $\braket{O}_\mathcal{A}^\text{b-mit}$ has smaller fidelity bias than the standard error-mitigated value $\braket{O}_\mathcal{A}^\text{QEM}$.
 To do so, we consider the simplified case of running digital ZNE \cite{Giurgica_Tiron_2020, 10313813, PhysRevX.7.021050, PhysRevLett.119.180509, Kandala2019, PhysRevA.102.012426} in the presence of uniform depolarizing noise across all two-qubit gates with error $p$, i.e.\ each two-qubit gate is followed by the two-qubit depolarizing channel
\begin{equation}
    \rho \mapsto (1-p) \rho + p \frac{\mathbb{I}}{4}.
\end{equation}
In this case it is possible to mitigate all error terms up to $\mathcal{O}(p^{n_\text{max}+1})$ using ZNE by taking a linear combination
\begin{align}
\begin{split}
    \braket{O}^\text{QEM} &= \sum_{n=0}^{n_\text{max}} \alpha(n) \braket{O}^\text{noisy}(2n+1) \\ &= \braket{O}^\text{exact} + \mathcal{O}(p^{n_\text{max}+1}),
    \label{eq:qem-interpolation}
\end{split}
\end{align}
where $\braket{O}^\text{noisy}(r)$ is the noisy expectation value of the circuit scaled by an odd-integer noise factor $r$, $n_\text{max}$ is the number of different noise amplifications used and the real coefficients $\alpha(n)$ are independent of the input circuit \cite{PhysRevA.102.012426}. The noise-scaled circuits are produced by replacing each two-qubit gate $U$ via\begin{equation}
    U \mapsto (UU^\dagger)^{(r-1)/2}U
\end{equation}
so that the circuits are functionally the same but contain $r$ times as many two-qubit gates.
Combining equations (\ref{eq:exp-val-fid}) and (\ref{eq:qem-interpolation}), the fidelity bias of an error-mitigated estimate is then
\begin{align}
\begin{split}
    b_{\mathcal{A}} &= \frac{\braket{O}^\text{QEM}}{\braket{O}^\text{exact}} - 1 \\
    &= \sum_{n=0}^{n_\text{max}} \alpha(n) \frac{\braket{O}^\text{noisy}(2n+1)}{\braket{O}^\text{exact}} - 1
    \label{eq:zne-bias}
\end{split}
\end{align}
which is of order $\mathcal{O}(p^{n_\text{max}+1})$. Now, suppose that our input application circuit contains $N$ two-qubit gates to which we assign the labels $1,2,\dots,N$. By considering the repeated action of the depolarizing channel, each noise-scaled noisy expectation value may be written \cite{PhysRevA.102.012426}
\begin{align}
    \begin{split}
  \braket{O}&^\text{noisy}(r) \\ 
  &=  (1-p)^{Nr}\braket{O}^\text{exact}\\
  &+(1-p)^{(N-1)r}(1-(1-p)^r) \sum_i \braket{O}^{\text{dep}(i)} \\
  &+(1-p)^{(N-2)r}(1-(1-p)^r)^2 \sum_{i_1, i_2} \braket{O}^{\text{dep}(i_1,i_2)} \\
  &+ \cdots \\
  &+ (1-(1-p)^r)^N \sum_{i_1, \dots, i_N} \braket{O}^{\text{dep}(i_1,\dots,i_N)},
  \label{eq:noisy-exp-val}
\end{split}
\end{align}
where $\braket{O}^{\text{dep}(i_1,\dots,i_m)}$ represents the exact expectation value for the evolution branch of the state in which the two-qubit gates with labels $i_1, \dots, i_m$ have been replaced by the two-qubit fully depolarizing noise channel $\rho \mapsto {\mathbb{I}}/{4}$. Thus, combining equations (\ref{eq:zne-bias}) and (\ref{eq:noisy-exp-val}), the fidelity bias may be written as a generic linear combination of these depolarizing expectation values,
\begin{equation}
    b_\mathcal{A} = \sum_{g=1}^N \ \beta(g)\sum_{i_1,\dots,i_g} \frac{\braket{O}_\mathcal{A}^{\text{dep}(i_1,\dots,i_g)}}{\braket{O}_\mathcal{A}^{\text{exact}}}
    \label{eq:application-bias}
\end{equation}
for some constants $\beta(g)$, and identically for $\mathcal{B}$ since the constants $\beta(g)$ are equal for both the application and benchmark circuits by construction. The bias of our bias-mitigated value given in equation (\ref{eq:bias-mit}) is then
\begin{align}
    \begin{split}
       & \  b_\mathcal{A} - b_\mathcal{B}  \\
       &= \sum_{g=1}^N \ \beta(g)\sum_{i_1,\dots,i_g} \left( \frac{\braket{O}_\mathcal{A}^{\text{dep}(i_1,\dots,i_g)}}{\braket{O}_\mathcal{A}^{\text{exact}}} - \frac{\braket{O}_\mathcal{B}^{\text{dep}(i_1,\dots,i_g)}}{\braket{O}_\mathcal{B}^{\text{exact}}}\right).
    \end{split}
    \label{eq:b-mit-bias}
\end{align}

We reiterate here that the right-hand side above is still a term of order $\mathcal{O}(p^{n_\text{max}+1})$. However, due to the identical structure of the application and benchmark circuits, we expect the bracketed terms in equation (\ref{eq:b-mit-bias}) to be similar in magnitude, resulting in a greatly reduced bias compared to the equivalent term in equation (\ref{eq:application-bias}). As we will see later, our numerical results (Figures \ref{fig:numerical-results} and \ref{fig:numerical-results-hardware}) strongly support this conclusion. 

\subsection{Benchmarked-noise zero-noise extrapolation (bnZNE)}
\label{sec:benchmarked-noise-zne}

Construction of these benchmark circuits allows us to define a modified zero-noise extrapolation method by improving the classical extrapolation procedure. As we will show later, we found this method to produce more accurate expectation values compared to standard ZNE. 

Recall that during digital ZNE (as opposed to pulse level modifications) of an application circuit $\mathcal{A}$, we extrapolate a series of datapoints $\left(r, \braket{O(r)}_\mathcal{A}\right)$ to $r=0$, where the $r\in R$ are the associated noise levels, calculated typically as the ratio between the number of two-qubit gates in the noise-inflated circuit versus the standard circuit (e.g. $R = \{1,3,5\}$). Here, we instead gauge the noise-level using our benchmark circuits; we hence call this method \textit{benchmarked-noise ZNE (bnZNE)}. 

Given as input an application circuit $\mathcal{A}$ and Pauli string observable $O$ acting non-trivially on some subset of qubits $Q \equiv \{i : O_i \neq I\}$, we define our method as follows:
\begin{enumerate}
    \item Define the benchmark circuit $\mathcal{B}(\mathcal{A})$. By condition C2, the measurement outcome of $\mathcal{B}$ is a single bitstring $b = (b_q)_{q \in Q}$.
    \item Run both $\mathcal{A}$ and $\mathcal{B}$ on the device at each of the chosen noise levels $r \in R$. For $\mathcal{A}$, calculate and store the noise-scaled expectation values $\braket{O(r)}_\mathcal{A}$. For $\mathcal{B}$, use the measurement outcomes to calculate the ensemble of probabilities $p(i;q,r)$ of obtaining measurement outcome $i \in \{0,1\}$ on qubit $q \in Q$ when running the noisy benchmark circuit with unitary $U_{\mathcal{B}(r)}$ at ZNE noise level $r \in R$,
    \begin{equation}
    \begin{split}
        p(i;q,r) &=  \text{Tr}_{\bar{q}}\left[\ket{i}\bra{i}_q U_{\mathcal{B}(r)}\rho_0U_{\mathcal{B}(r)}^{\dagger}\right] \\
        &= \sum_j \bra{j}_{\bar{q}} \bra{i}_q \left( U_{\mathcal{B}(r)}\rho_0U_{\mathcal{B}(r)}^{\dagger} \right) \ket{i}_q \ket{j}_{\bar{q}},
    \end{split}
    \end{equation}
    where $\bar{q}$ denotes all qubits except for $q$ and $\rho_0$ is the initial all-zero state.
    \item Calculate the associated error rate $\varepsilon(r)$ at ZNE noise level $r$, which we define to be
    \begin{equation}
        \varepsilon(r) \equiv \prod_{q \in \mathcal{Q}} (1-p(b_q; q,r)),
        \label{eq:bnzne-error-rate}
    \end{equation}
    i.e.\ the product of the probabilities of obtaining the wrong measurement outcome across each of the measured qubits. 
    \item Extrapolate the datapoints $\left(\varepsilon(r), \braket{O(r)}_\mathcal{A}\right)$ to noise level $\varepsilon = 0$ to reconstruct the corresponding noiseless value.
\end{enumerate}

As we will show later, we found our method to produce more accurate datapoints when compared to standard ZNE. 
We note however that this method requires twice as many circuit runs compared to regular ZNE, since for each noise-scaled application circuit we must also run the corresponding benchmark circuit.

This method is inspired by other recent methods for modified zero-noise extrapolation which seek to improve the extrapolation procedure by improving the noise level calculation \cite{PhysRevA.110.042625, kim2025enhancedextrapolationbasedquantumerror}. One method, \textit{inverted-circuit ZNE (IC-ZNE)} \cite{PhysRevA.110.042625},  gauges the noise-level by applying the noise-scaled circuit followed by its inverse, and measuring the probability of the all-zero state. Another considers circuits with repeating layers and benchmarks the fidelity of a single-layer \cite{kim2025enhancedextrapolationbasedquantumerror}. We believe our method is advantageous in that it requires no overhead of circuit depth compared to the original circuit beyond standard ZNE and is hence more scalable to large-scale NISQ experiments. It also makes no assumptions on the circuit structure. In addition, we consider the probability of measurement outcomes for individual qubits -- which tends to $\frac{1}{2}$ as the state becomes depolarized -- rather than the probability of measuring the all-zero output state used by both works above, which decays to zero exponentially with both the number of qubits and circuit depth. Moreover, we demonstrate our work using 100-qubit experiments compared to the at most 4 and 6 qubits used in these two works respectively.

In the appendix \ref{sec:ic-zne}, we examine the relationship between bnZNE and IC-ZNE more closely. We observe that IC-ZNE breaks down for utility scale circuits with 100 qubits, most likely due to the exponentially decaying all-zero probability mentioned above. To that end, we improve the IC-ZNE results by using a scalable error rate calculation as in equation (\ref{eq:bnzne-error-rate}). This, however, also breaks down more quickly than bnZNE for large circuit depths, likely since the IC-ZNE method uses a benchmark circuit which is twice as deep as the application circuit. 

\section{Generating benchmark circuits}
\label{sec:generating-benchmark-circuits}

Given an application circuit and target hardware platform as input, we present two methods for generating the corresponding benchmark circuits. The first is intentionally hardware agnostic, meaning it can be applied to any gate-based hardware platform. The second is designed to be tailored to a specific hardware's native gate set and we use the IBM superconducting devices as an example. We discuss the potential limitations of these two methods in section \ref{sec:discussion}. Generally, however, benchmark circuits need only satisfy the criteria C1--3 stated at the beginning of section \ref{sec:bias-mitigation-using-benchmark-circuits}. In appendix \ref{sec:entangling-benchmarks-for-trotterized-circuits}, we give an additional method which is tailored to layered applications with invertible layers, such as Trotterized Hamiltonian simulations.

\subsection{Hardware-agnostic benchmark circuits}
\label{sec:hardware-agnostic}

To generate the hardware-agnostic benchmark circuits, we assume our input application circuit to be expressed in terms of various single and two-qubit Pauli rotations which have broadly similar error rates. To apply this method to realistic hardware, it will then be necessary to transpile each Pauli rotation to the native gate set with comparable gate error. This makes our method highly transferable to different hardware types, typically at the expense of an overhead of single-qubit gates when compared to the most efficient transpilation of the application circuit. We discuss this more later.

We begin with some useful definitions. We define an \textit{$n$-qubit Pauli rotation gate} as $R_P(\theta) = e^{-i\theta P/2}$, where $P \in \{I,X,Y,Z\}^{\otimes n}$ is a Pauli string of length $n$. We also define $T(P) = \{i \ |  P_i \neq I\}$ to be the set of qubits on which $P$ acts non-trivially, e.g. $T(IIXIZ) = \{3,5\}$, and the \textit{weight} $w(P) = |T(P)|$ to be the number of such qubits. For $P \in \{X,Y,Z\}$ we also define $\ket{P^{\pm 1}}$ to be the single-qubit eigenstate of $P$ with eigenvalue $\pm1$, so that for example $\ket{Z^{+1}} = \ket{0}$ and $\ket{X^{-1}} = \ket{-}$. 

The set of all single and two-qubit Pauli rotations is universal for quantum computing \cite{NandC}; hence we consider for simplicity our $n$-qubit input application circuit $\mathcal{A}$ to be decomposed in terms of $M$ such rotations,
\begin{equation}
    \mathcal{A} = R_{P_M}(\theta_M) \cdots R_{P_1}(\theta_1),
\end{equation}
where each $P_i$ has length $n$ and weight 1 or 2. Our work can be straightforwardly extended to larger multi-qubit gates if needed, for example in ion-trap hardware where native gates can include multi-qubit interactions \cite{PhysRevLett.87.257904, Johanning_2009, Bassler2023synthesisof, PhysRevLett.117.220501}.

Suppose that $\mathcal{A}$ is an $n$-qubit circuit with input state $\ket{0}^{\otimes n}$. To generate a single Clifford benchmark circuit $\mathcal{B}$, we first start with the ansatz 
\begin{equation}
    \mathcal{B}_0 = R_{\tilde{P}_M}(\tilde{\theta}_M) \cdots R_{\tilde{P}_1}(\tilde{\theta}_1),
\end{equation}
and pick the Paulis $\tilde{P}_i$ and angles $\tilde{\theta}_i$ via the following algorithm:
\begin{enumerate}
    \item Let $\ket{\psi_{q}}$ denote the state of qubit $q$ and set $\ket{\psi_{q}} = \ket{Z^{+}}$ for all $q = 1, \dots, n$ as our input state. We will update these as we move through the circuit to keep track of the overall circuit state, which will remain a product state throughout. 
    \item For $i = 1, \dots, M$:
    \begin{enumerate}
        \item If $w(P_i) = 1$, pick uniformly at random $\tilde{\theta}_i \in \{\pi/2, \pi, 3\pi/2\}$ and $\tilde{P}_i \in \{X,Y,Z\}$. Let $T(P_1) = \{a\}$ and update $\ket{\psi_{a}} \mapsto  R_{\tilde{P}_i}(\tilde{\theta}_i) \ket{\psi_{a}}$;  this will produce another Pauli eigenstate.
        \item If $w(P_i) = 2$, set $\tilde{\theta}_i= \pi$. Let the elements of $T(P_i)$ be $a,b$ in a random order. Write the current state of qubit $a$ as $\ket{\psi_{a}} = \ket{Q_a^{c_a}}$ where $Q_a \in \{X,Y,Z\}$ and $c_a = \pm 1$, and likewise for qubit $b$. Set $\tilde{P}_a = Q_a$ and pick $\tilde{P}_b \in \{X,Y,Z\}$ uniformly at random. If $\tilde{P}_b \neq Q_b$, update $\ket{\psi_{b}} \mapsto \ket{Q_b^{-c_b}}$.
    \end{enumerate}
\end{enumerate}

The output state of this sequence of rotations is a known product state of Pauli eigenstates, $\otimes_{q}\ket{\psi_q}$. We can hence rotate each qubit to be a +1-eigenstate of the Pauli observable $O = \otimes_{q=1}^n O_q$ by applying a single `correction' layer of single-qubit Pauli rotation gates:
\begin{equation}
    \prod_{q \in T(O)} R_{C_q}(\phi_q)
    \label{eq:correction-layer}
\end{equation}
where $C_q \in \{I,X,Y,Z\}$ and $\phi_q \in \{0, \pi/2, \pi,  3\pi/2\}$ are chosen such that $R_{C_q}(\phi_q) \ket{\psi_q} = \ket{O_q^{+1}}$, the +1 eigenstate of $O_q$. The full benchmark circuit is then
\begin{equation}
    \mathcal{B} = \left(\prod_{q \in T(O)} R_{C_q}(\phi_q)\right) R_{\tilde{P}_M}(\tilde{\theta}_M) \cdots R_{\tilde{P}_1}(\tilde{\theta}_1).
\end{equation}

This procedure results in a Clifford circuit $\mathcal{B}$ which, compared to our original application circuit $\mathcal{A}$, has the same number of Pauli rotation gates, acting in the same places on the same qubits, plus an additional depth-1 correction layer of single-qubit Pauli rotations at the end of the circuit. Moreover, $\mathcal{B}$ has $\braket{O}_\mathcal{B}^\text{exact} = 1$.

To satisfy the condition C1 that $\mathcal{A}$ and $\mathcal{B}$ will have identical gate structure upon compilation (see section \ref{sec:bias-mitigation-using-benchmark-circuits}), we add a trivial layer of single-qubit Pauli rotations to the end of the application circuit, so that the final application circuit is
\begin{equation}
    \mathcal{A} = \left(\prod_{q \in T(O)} R_{D_q}(2\pi)\right) R_{{P}_M}({\theta}_M) \cdots R_{{P}_1}({\theta}_1)
\end{equation}
where the $D_q \in \{I,X,Y,Z\}$ are chosen uniformly at random.

We can then estimate the noisy expectation value fidelity distribution of our application circuit by simply taking measurements of $O$ with respect to our benchmark circuits. This allows us to benchmark the bias and variance that we would expect across circuit runs, both with and without error mitigation.

For these benchmark circuits to be able to accurately benchmark the performance of our application, we must assume that the gates $R_{P_i}(\theta_i)$ and $R_{\tilde{P}_i}(\tilde{\theta}_i)$ are subjected to very similar errors with weak angle-dependence when transpiled and and implemented on the target hardware. This assumption holds, for example, for any hardware platform in which
\begin{itemize}
    \item the native single-qubit [two-qubit] gates have very similar error rates;
    \item any single-qubit [two-qubit] Pauli rotation may be transpiled using the same number of single and two-qubit gates in the same order.
\end{itemize}
In our first numerical simulations (Figure \ref{fig:numerical-results}), we satisfy these constraints by introducing \textit{rigid transpilation rules} for transpiling Pauli rotations for the IBM superconducting hardware such that all Pauli rotations use the same native gate structure; we give more details appendix \ref{app A}. This introduces a multiplicative overhead of single-qubit gates compared to the most efficient transpilation of our application. In the next section, we address this overhead using knowledge of the target hardware, reducing it to zero for the case of recent IBM devices. 

\subsection{Hardware-tailored benchmark circuits}
\label{sec:hardware-tailored-benchmarking-circs}

The Pauli rotations approach of the previous section allows us to produce very general, random benchmark circuits. However, one downside is that consistent transpilation of all the circuits to a specific hardware native gate set -- such that all circuits contain the same types of gates in the same places -- will typically require an overhead of single qubit gates, degrading circuit quality compared to the most gate-efficient transpilation possible for the application. This is particularly the case when the native two qubit gate is not a Pauli rotation. In this section, we give sufficient conditions such that it is possible to construct benchmark circuits directly from the hardware-transpiled application circuit, thus dramatically reducing our overhead compared to the hardware-agnostic approach. We show that this method can be applied straightforwardly to the IBM \textit{Heron r2} architecture.

We firstly assume that the hardware's native gate set contains one two-qubit gate; the generalisation to many such gates is straightforward. A set of general, sufficient conditions on the hardware for the ability to generate hardware-tailored benchmark circuits is as follows. 
\begin{enumerate}
    \item There exists an orthogonal single-qubit basis set $\{\ket{\psi_1}, \ket{\psi_2}\}$ such that the native two-qubit gate maps tensor products of these states to tensor products, i.e. for all $i,j \in \{1,2\}$ there exist $k, l \in \{1,2\}$ such that $\ket{\psi_i}\otimes\ket{\psi_j} \mapsto \ket{\psi_k}\otimes\ket{\psi_l}$ under the native two-qubit gate.
    \item Each single-qubit gate can be replaced with another single-qubit gate of equivalent error (e.g.\ by changing an angle parameter or replacing the gate entirely) such that it acts on the states $\{\ket{\psi_1}, \ket{\psi_2}\}$ either as identity or exchanging $\ket{\psi_1} \leftrightarrow \ket{\psi_2}$.
\end{enumerate}
To generate a benchmark circuit, we then apply the following procedure:
\begin{enumerate}
    \item Map the input $\ket{00\cdots0}$ into the $\{\ket{\psi_1}, \ket{\psi_2}\}$ basis; this can in general be done using a small number of single-qubit gates per qubit \cite{PhysRevA.52.3457}.
    \item Apply the application circuit with gate replacements according to point 2 above.
    \item For each qubit $q$ on which $O$ acts non-trivially, map the state $\ket{\psi_i}_q \mapsto \ket{O_q^{+1}}_q$, where we adopt the notation of the previous section.
\end{enumerate}

At most, this introduces a small additive (rather than multiplicative as before) overhead of single-qubit gates at the beginning and end of the circuit. Notably, the resulting circuit need not be Clifford but the expectation value is efficiently classically tractable. 

We consider as an example the most recent \textit{Heron r2} IBM superconducting quantum computers.
These devices use $CZ$ as the native two-qubit gate, $R_z(\theta), \sqrt{X}, X$ as the native single-qubit gates and measure in the standard $Z$-basis. In particular, the $\sqrt{X}$ and $X$ gates are implemented via equal-time pulses, such that they are reported to have identical error rates \cite{ibm_quantum}. 
The $R_z$ gate is simulated virtually for arbitrary angle by adjusting the relative phase between the projection of the qubit vector on the Bloch sphere and the microwave control tone and can be viewed as error-free \cite{PhysRevA.96.022330}, although this is not a constraint.
We note that the $CZ, R_Z(\theta)$ and $X$ gates act trivially on tensor products of the standard $\{\ket{0},\ket{1}\}$ basis -- mapping tensor products to tensor products while introducing at most an irrelevant global phase. Hence, we generate benchmark circuits from the application circuits as follows: simply replace $\sqrt{X}$ gates with $X$ gates. The resulting output state is an efficiently computable classical bitstring such that any noiseless expectation values $\braket{O} = \pm 1$ where $O \in \{I,Z\}^{\otimes n}$ as discussed above. We can assume that $\braket{O} = +1$, else one may apply a classical bit flip to the measured output state. This method of generating benchmark circuits introduces no gate overhead into the benchmark circuits, using as few gates as the most efficient transpilation of the application circuit. 

One natural downside to both of the methods introduced in this section is that the benchmark circuits do not introduce entanglement, and thus are less prone to long-range crosstalk errors that are typical of many real devices. We discuss this further in section \ref{sec:discussion} and give a further method for generating entangling benchmark circuits in appendix \ref{sec:entangling-benchmarks-for-trotterized-circuits}.


\section{Results}
\label{sec:numerical-results}

We demonstrate our methods both numerically and experimentally. First, we directly verify our analytical results from section \ref{bias-mit-subsec} by benchmarking zero-noise extrapolation in a depolarizing noise model using small-scale classical simulation and our hardware-agnostic method of benchmark circuit generation (section \ref{sec:hardware-agnostic}). To do this, we consider two quantum simulation applications, since these have straightforward expressions in terms of Pauli rotations.

Second, we employ our hardware-tailored benchmark circuits (section \ref{sec:hardware-tailored-benchmarking-circs}) by running utility-scale 100-qubit experiments at varying circuit depths to compare results from regular zero-noise extrapolation (ZNE), bias-mitigated ZNE (b-mit ZNE), benchmarked-noise ZNE (bnZNE) and bias-mitigated bnZNE (b-mit bnZNE). We compare our results against the exact values obtained using classical tensor network simulations. Additional further hardware results are given in the appendix.

In the associated GitHub repository \cite{Github}, we provide software for implementing each of the error mitigation methods used in this work, as well as the obtained and post-processed hardware data. We give a brief overview of the software in appendix \ref{app:software}.

\subsection{Numerical Results}

We first run classical simulations with the goal of empirically validating the result of section \ref{bias-mit-subsec}, namely that our method reduces the bias of zero-noise extrapolation under depolarizing noise. To do so, we consider two applications. The first is the evolution of a two-dimensional kicked Ising model with initial state $\ket{0}^{\otimes{N}}$ and Hamiltonian
\begin{equation}
    H_1 = -J\sum_{\braket{i,j}}Z_i Z_j + h\sum_i X_i,
\end{equation}
where $J$ is the coupling strength of nearest-neighbour spins and $h$ is a global transverse field. This application was recently used by IBM and others to claim demonstrations of the utility of quantum hardware \cite{Kim2023, PRXQuantum.5.010308, doi:10.1126/sciadv.adk4321, KECHEDZHI2024431, anand2023classical, shao2023simulating, liao2023simulation, rudolph2023}. The first sum runs over all pairs of nearest-neighbour qubits $(i,j)$ and the second runs over all qubits in the topology, which we take for each plot to be a 10-qubit two-dimensional subset of the hardware layout. Specifically, we use the heavy-hexagon layout of the recent IBM superconducting devices \cite{ibm_quantum}. We measure the magnetization of a single site $i$ in the center of the topology, $O = Z_i$. Our application circuit is then the Trotterized time evolution of this Hamiltonian,
\begin{equation}
    \mathcal{A} = [L_2L_1]^{N_T},
    \label{eq:ki-circ}
\end{equation}
applied to the $\ket{0}^{\otimes n}$ state, where
\begin{equation}
    L_1 = \prod_{i} R_{X_i} \left(\theta_1 \right),  \quad  L_2 = \prod_{\braket{i,j}} R_{Z_i Z_j}\left( \theta_2 \right).
\end{equation}
For simplicity we set $\theta_1 = \theta_2 = 0.01$ for our classical simulations and vary only the number of Trotter layers ${N_T}$. The second application is time evolution of the quantum Heisenberg model \cite{arovas1988functional} with initial state $\ket{0}^{\otimes N}$ and Hamiltonian
\begin{equation}
    H_2 = -J\sum_{\braket{i,j}}(X_i X_j + Y_iY_j + Z_iZ_j) - h \sum_i (X_i+Z_i),
\end{equation}
again applied to a 10-qubit connected subset of the hardware topology as above. The application circuit in this case is
\begin{equation}
    \mathcal{A} = [L_4L_3]^{N_T}
\end{equation}
where
\begin{align}
    L_3 &= \prod_{i} R_{X_i}(\theta_3)R_{Z_i} (\theta_3), \\  L_4 &= \prod_{\braket{i,j}} R_{X_i X_j}\left( \theta_{4} \right)R_{Y_i Y_j}\left( \theta_{4} \right)R_{Z_i Z_j}\left( \theta_{4} \right)
\end{align}
and we set $\theta_3 = \theta_4 = 0.01$ for our classical simulations and vary only ${N_T}$.

We transpile these circuits to the native gate set $\{CZ, R_Z(\theta), X, \sqrt{X}\}$ of the Heron r2 architecture using the hardware-agnostic benchmark circuits described in section \ref{sec:hardware-agnostic}. This requires using a consistent transpilation strategy such that each single [two]-qubit Pauli rotation is transpiled into the same native gate structure; more details can be found in the appendix \ref{app A} or in the attached GitHub repository \cite{Github}.

For both applications, we pick small rotation angles for our small-scale classical simulations in order to mirror the usual structure of Trotter circuits. We consider more generic angles later in our large-scale hardware simulations.

For error mitigation, we employ digital ZNE combined with Pauli twirling \cite{WallmanRandomizedCompiling, Ware_2021, Kim2023}. Although Pauli twirling is not strictly necessary here since we simulate only depolarizing noise, this allows us to consider running equivalent numbers of modified circuits as to what would be required on hardware. For ZNE, we locally fold the two-qubit $CZ$ gates at noise levels $r=1,3,5$ by replacing each two-qubit gate $CZ \mapsto (CZ)^r$ and use 5 twirled circuits per noise level. Folding only the two-qubit gates produces an error mitigation bias which depends on the number of Trotter layers. The expectation values $\braket{O}(r)$ at each noise level $r$ are fitted to an exponential function and the $r=0$ value is extrapolated. In the benchmark case we run 5 benchmark circuits per datapoint and plot the $3\sigma$ errorbars. This allows us to capture both the bias and variance of the error mitigation method. 

\begin{figure}[]
    \hspace*{-1em}
    \includegraphics[scale=0.725]{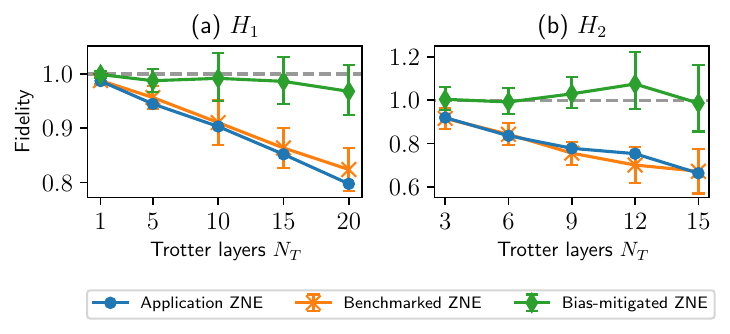}
    \caption{
    Bias mitigation of zero-noise extrapolation using 10-qubit simulations of the two-dimensional kicked Ising ($H_1$) and Heisenberg ($H_2$) models with depolarizing noise. We benchmark both the bias and variance of the application error mitigation (circles) using the benchmark circuits (crosses) with $3\sigma$ errorbars, then mitigate this bias (diamonds). In all cases we see almost full elimination of the fidelity bias, validating our analytical results from section \ref{bias-mit-subsec}.
     }
    \label{fig:numerical-results}
\end{figure}

We present our numerical results in Figure \ref{fig:numerical-results}, where we plot the number of Trotter layers against the expectation value fidelity (equation (\ref{eq:exp-val-fid})) such that a value of 1 indicates perfect error mitigation. In both cases we see near-perfect mitigation of the bias (diamonds) compared to the original error mitigation data (circles), demonstrating the mitigated bias (equation (\ref{eq:b-mit-bias})) to be significantly smaller than the unmitigated bias (equation (\ref{eq:application-bias})). In both cases, the benchmarked error mitigation fidelities (crosses) accurately mirror those of the application circuit. This allows us to quantify both the bias and variance of the error-mitigated expectation value fidelity distribution. 

\subsection{Experimental Results}
\label{sec:experimental-results}
Next, we demonstrate the resilience of our methods to real noise in utility-scale experiments by performing the Hamiltonian simulation of a 100-qubit one-dimensional kicked Ising chain (equation (\ref{eq:ki-circ})) on the 156-qubit \texttt{ibm\textunderscore fez} device, considering both single and two-qubit derived observables. To do so, we employ the hardware-tailored benchmark circuits introduced in section \ref{sec:hardware-tailored-benchmarking-circs}.

\begin{figure}[]
    \hspace*{-1em}
    \includegraphics[scale=0.75]{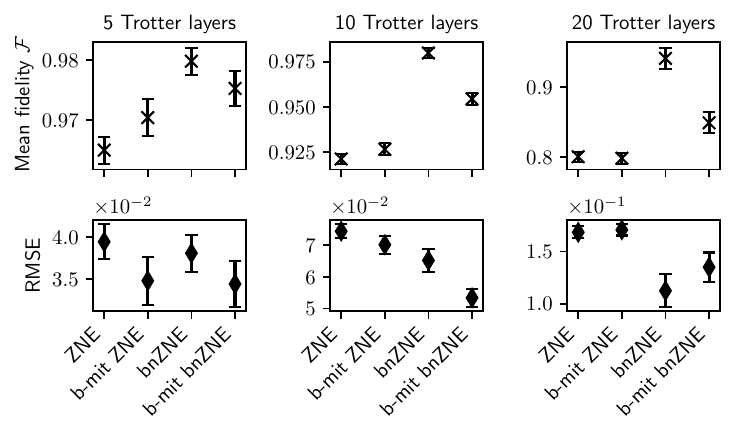}
    \caption{
     Measurement results of single-site $\braket{Z}$ expectation values from 100-qubit experiments using the \texttt{ibm\textunderscore fez} superconducting device. We propagate a one-dimensional 100-site kicked Ising chain and perform (a) ZNE, (b) bias-mitigated ZNE, (c) bnZNE, and (d) bias-mitigated bnZNE. We also use Pauli twirling, DD and TREX, see text for details. We measure single-site magnetisations $\braket{Z}$ for each qubit and plot the mean fidelity averaged over all 100 qubits with $3\sigma$ errorbars. We also plot the RMSE calculated over all sites. Circuits with 5, 10, 20 Trotter layers contained 495, 990, 1980 $\text{CZ}$ gates respectively. We see in each case that our bnZNE method outperforms standard ZNE in terms of both the mean fidelity and RMSE. For 5 and 10 Trotter layers, we see an improvement to our results via the bias mitigation.
     }
    \label{fig:numerical-results-hardware}
\end{figure}

Firstly, we consider measurement of the single-site magnetisations $\braket{Z_i}$. In Figure \ref{fig:numerical-results-hardware}, we plot the mean measured expectation value fidelity averaged over all N=100 qubits,
\begin{equation}
    \mathcal{F} \equiv \frac{1}{N} \sum_{q=1}^{N} \frac{\braket{Z_q}_\mathcal{A}^\text{QEM}}{\braket{Z_q}_\mathcal{A}^\text{exact}}
\end{equation}
and the root mean squared error (RMSE) 
\begin{equation}
    \text{RMSE} \equiv \sqrt{\frac{1}{N} \sum_{q=1}^{N} \left( \braket{Z_q}_\mathcal{A}^\text{QEM} - \braket{Z_q}_\mathcal{A}^\text{exact} \right)^2 }
\end{equation}

for each of the QEM methods (a) ZNE, (b) bias-mitigated ZNE, (c) bnZNE (section \ref{sec:benchmarked-noise-zne}), and (d) bias-mitigated bnZNE. To improve noise resilience, for each of the QEM methods we also perform 32 instances of Pauli twirling \cite{WallmanRandomizedCompiling, Ware_2021, Kim2023} per circuit, per ZNE level,  as well as dynamical decoupling (DD) \cite{Ezzell2023} and twirled readout error extinction (TREX) \cite{PhysRevA.105.032620}; see appendix \ref{app EM} for more details. Each circuit instance was run with 2048 shots. We pick non-trivial angles $\theta_1 = -\pi/8, \  \theta_2 = -\pi/2$ in order to generate large-magnitude expectation values to reduce the effects of shot noise. The exact expectation values for the application circuits were calculated using classical matrix product state simulations with ITensors.jl \cite{itensor}.

We see in each case that our bnZNE method -- both with and without bias mitigation -- outperforms standard ZNE in terms of both the mean fidelity and RMSE, indicating not only an improvement in bias but a reduced variance. Interestingly, bias-mitigation tends to reduce the accuracy of bnZNE despite often improving the RMSE, i.e. decreasing the variance. In fact, within the $3\sigma$ uncertainty levels reported here, bias mitigated methods consistently show a significant improvement in variance over standard ZNE, making these potentially more reliable when being applied to circuits whose exact value is unknown. We emphasise here that each of the methods (b)-(d) uses precisely twice as many circuit runs as for regular ZNE (a), and that the same runs can be used for each method.

\begin{figure}
    \centering
    \includegraphics[scale=0.75]{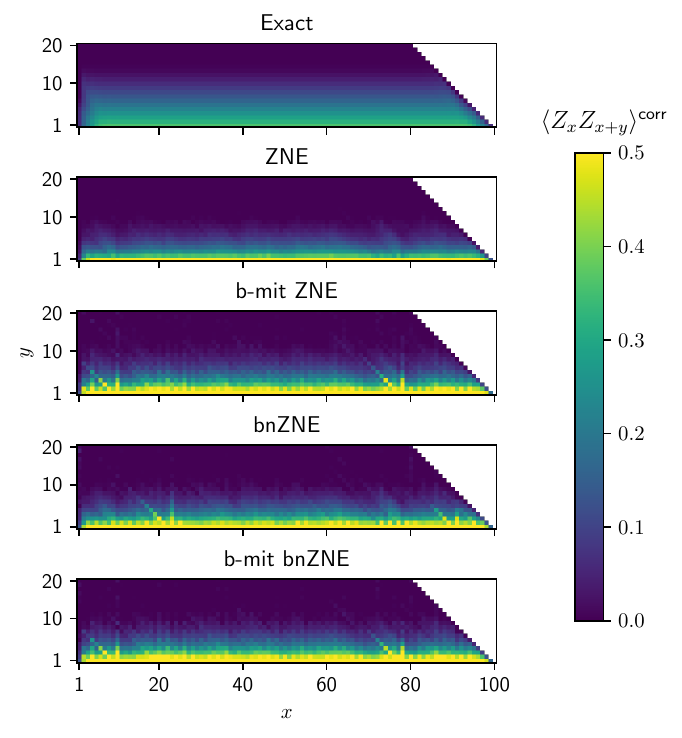}
    \caption{Measurement of the two-qubit correlations $\braket{Z_xZ_{x+y}}^\text{corr}$ for a 100-site one-dimensional kicked Ising chain with 20 Trotter layers ran on the \texttt{ibm\textunderscore fez} superconducting device. We compare the exact values against results obtained using the ZNE, bias-mitigated ZNE, bnZNE, and bias-mitigated bnZNE error mitigation methods.}
    \label{fig:correlations-unentangled}
\end{figure}

Next, we consider the measurement of multi-qubit observables. Specifically, we will consider the measurement of two-qubit correlations between qubits $i$ and $j$ in the Ising chain,
\begin{equation}
    \braket{Z_i Z_j}^\text{corr} \equiv \braket{Z_i Z_j} - \braket{Z_i}\braket{Z_j}
\end{equation}
and their decay rates. In Figure \ref{fig:correlations-unentangled} we run the 100-qubit kicked Ising circuit with 20 Trotter layers and measure the correlations between each of the 100 sites $x$ and those a distance $y$ further along the chain in a fixed direction. We see in particular that our three methods are able to recover long-range correlations at greater distances than regular ZNE, although close-range values appear overestimated. We are also clearly able to identify problematic qubits in the chain in the two bias-mitigated plots; we discuss this further in appendix \ref{sec:identification-of-hardware-imperfections}. 

\begin{figure}
    \centering
    \includegraphics[scale=0.75]{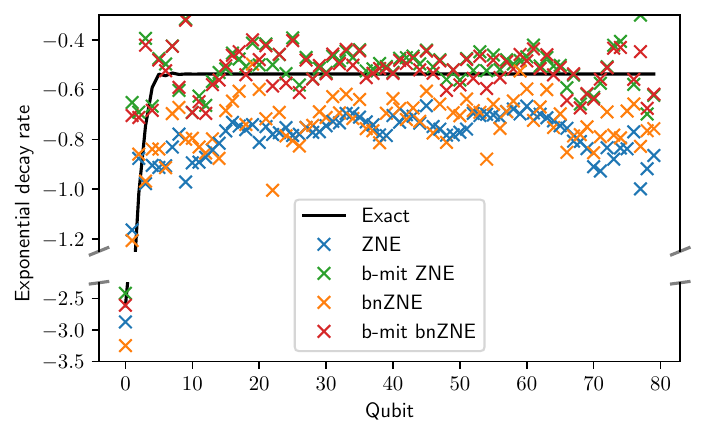}
    \caption{Measurement of the exponential decay rate in two-qubit correlations for the first 80 qubits in the 100-qubit chain. For each qubit $q$ we measure the exponentially decaying correlations $\braket{Z_q Z_{q+i}}^\text{corr}$ for $i=1,2,\dots,20$ using each QEM method and extract the decay rate. We observe that, whilst our bnZNE still outperforms standard ZNE, the two bias-mitigated methods offer significant improvements with b-mit bnZNE performing the best.}
    \label{fig:decay-rates-unentangled}
\end{figure}

As an example of a more complicated physical parameter, we also measure the exponential decay rates of these correlations with distance for the first 80 qubits in the chain. That is, we use the relation $\braket{Z_x Z_{x+y}}^\text{corr} \sim e^{-\alpha_x y}$ \cite{Lieb1972, PhysRevLett.97.050401, doi:10.1142/9789814415255_0002} and extract the decay rates $\alpha_x$ for $x=1,2,\dots,80$ and $y = 1,2,\dots,20$. We plot the results in Figure \ref{fig:decay-rates-unentangled}. In this case, we see a clear improvement over ZNE for all three of our methods in estimating the decay rates. In particular, the bias-mitigation makes a significant improvement over its biased counterpart for both ZNE and bnZNE, with the latter performing best. We deduce therefore that our bias-mitigation method can be used to make a significant improvement in the measurement of physically relevant quantities derived from correlated expectation values. We also note that in this case, our bias mitigation and bnZNE come at the expense of increased variance versus regular ZNE, highlighting the bias-variance trade-off that is more generally observed in quantum error mitigation methods \cite{Endo2021}.  

\section{Discussion}
\label{sec:discussion}
In this work, we presented a method to benchmark the bias and variance of any biased error mitigation protocol. We also introduced benchmarked-noise zero-noise extrapolation (bnZNE) as an adaptation of ZNE. In terms of the circuit run overhead, each benchmark circuit requires only as many device runs as would be used for standard error mitigation of the application circuit, thus making the procedure relatively low-cost: for a given error mitigation method requiring $N$ circuit runs, the cost of the bias-mitigated method is simply $2N$. We gave analytical evidence that for depolarizing noise and ZNE our method will result in a bias reduction, and have verified this fact using classical simulation. We also performed 100-qubit experiments on the IBM superconducting hardware to show that our methods can be used to boost the quality of results for utility-scale applications under real-world noise. We provided two methods for generating benchmark circuits, although in general there may be arbitrarily many possibilities for benchmark circuits which obey the conditions set out in section \ref{sec:bias-mitigation-using-benchmark-circuits}.

We note that there also exist unbiased methods for error mitigation such as probabilistic error cancellation \cite{vandenBerg2023} or tensor error mitigation \cite{filippov2023scalabletensornetworkerrormitigation,Fischer2026}. However, these methods comes at the cost of a sampling overhead which scales exponentially with the average number of faults in each circuit run and extremely sophisticated classical pre- and post-processing \cite{RevModPhys.95.045005,filippov2024scalabilityquantumerrormitigation}. Our work has the potential to enable high-quality results using lower-cost biased methods instead. 

We conclude by discussing the benchmark circuits. In this work, we intentionally provided a method of generating benchmark circuits which is agnostic to the both the target application circuit and hardware. This approach comes with a cost in circuit depth, typically introducing additional single-qubit gates since we rely on being able to consistently transpile each single or two-qubit Pauli rotation using the same native gate structure. We also provided a specific example of how to generate efficient circuits for the IBM hardware which introduced no gate overhead compared to the most efficient transpilation of the standard application circuit. However, we reemphasize that our work is entirely modular, and these methods for generating benchmark circuits can be straightforwardly replaced with other methods which may be more gate-efficient for certain applications or architectures. One downside to our two examples of benchmark circuits, for example, is that they do not introduce entanglement -- the circuit state remains a product state throughout. Hence, in practice they are potentially more noise-resilient than the application circuits and hence do not fully capture the effect of crosstalk noise on the application circuit. We see this in Figure \ref{fig:numerical-results-hardware} where in many cases the bias is largely reduced but not entirely eliminated. In the appendix \ref{sec:entangling-benchmarks-for-trotterized-circuits}, we give a method for implementing entangling benchmark circuits tailored to applications with a layered structure such as Trotterized Hamiltonian simulations, and provide additional hardware results for this method. This resulted in greater overall improvements when compared to the fidelities obtained via standard ZNE, although we observed the overall quality of data obtained from the hardware to be worse in this case, possibly due to an unfortunately-timed execution within the calibration window. 

Another approach to benchmarking QEM methods which has been considered before is to simply `Cliffordize' the circuit, for example in the IBM case by setting each $\theta$ to be a multiple of $\frac{\pi}{2}$ in each $R_Z(\theta)$ gate to produce a circuit whose evolution can be simulated classically \cite{osti_319738}. However, the logical function of such a high entangling circuit is not easily controllable and thus one cannot necessarily extract meaningful (non-zero) expectation values. Recent work has considered the relationship between the fidelities of application circuits and their Clifford counterparts \cite{merkel2025cliffordbenchmarkssufficientestimating}, but has not as of yet considered the effects of error mitigation. In theory, such methods -- if successful -- could be combined with this work. 

Hence, future research should try to identify classes of benchmark circuit which accurately capture the effects of noise on the application through both gate structure and entanglement. In addition, we should identify situations -- combinations of error mitigation method, application and hardware -- where highly tailored benchmark circuits can be produced.

\section{Acknowledgments; data and code availability}
We thank Benedikt Fauseweh, Michael Epping, and Frank Wilhelm-Mauch for their suggestions and fruitful discussion over the course of this work. This research was made possible by the DLR Quantum Computing Initiative \cite{ALQU} and the German Federal Ministry of Research, Technology and Space \cite{BMFTR}.

The program code and data produced during this work can be found in the associated GitHub repository \cite{Github}.

We acknowledge the use of IBM Quantum services for this work. The views expressed are those of the authors, and do not reflect the official policy or position of IBM or the IBM Quantum team. All quantum circuits were produced using Qiskit \cite{qiskit2024}.

%



\appendix

\begin{table*}[]
\begin{tabular}{l|l|}
\cline{2-2}
                                                      & \textbf{Customization options}                                                                                                                                                   \\ \hline
\multicolumn{1}{|l|}{Application circuit}                     & - Input application circuit                                                                                                                                                      \\ \hline
\multicolumn{1}{|l|}{Benchmarking circuit generation} & \begin{tabular}[c]{@{}l@{}}- Generation method: hardware-agnostic (Pauli rotations), IBM hardware-tailored \\ or entangling benchmarks for Trotterized applications\end{tabular} \\ \hline
\multicolumn{1}{|l|}{ZNE with bias-mitigation}        & \begin{tabular}[c]{@{}l@{}}- On/Off \\ - Fitting method (exponential, linear, polynomial)\\ - Noise levels (e.g. {[}1,3,5{]})\end{tabular}                                       \\ \hline
\multicolumn{1}{|l|}{bnZNE with bias-mitigation}      & \begin{tabular}[c]{@{}l@{}}- On/Off\\ - Fitting method\\ - Noise levels\end{tabular}                                                                                             \\ \hline
\multicolumn{1}{|l|}{IC-ZNE with bias-mitigation}     & \begin{tabular}[c]{@{}l@{}}- On/Off\\ - Fitting method\\ - Noise levels\end{tabular}                                                                                             \\ \hline
\multicolumn{1}{|l|}{PT}                              & \begin{tabular}[c]{@{}l@{}}- On/Off\\ - Number of twirled instances\\ - Average over instances before/after extrapolation\end{tabular}                                           \\ \hline
\multicolumn{1}{|l|}{DD}                              & - On/Off                                                                                                                                                                         \\ \hline
\multicolumn{1}{|l|}{TREX}                            & - On/Off                                                                                                                                                                         \\ \hline
\multicolumn{1}{|l|}{IBM backend}                     & \begin{tabular}[c]{@{}l@{}}- Input (real or simulated) backend\\ - Number of shots per circuit\\ - Observables to be measured\end{tabular}                                       \\ \hline
\end{tabular}
\caption{Overview of customization options included in the attached software \cite{Github}. We support bias-mitigation of zero-noise extrapolation (ZNE), inverse-circuit ZNE (IC-ZNE) and our introduced benchmarked-noise ZNE (bnZNE) method. To improve resilience to hardware noise, these can be combined with Pauli twirling (PT), dynamical decoupling (DD) and twirled readout error extinction (TREX) which are each described in appendix \ref{app EM}.}
\label{table:qem}
\end{table*}

\section{Overview of error mitigation methods used}\label{app EM}
Here, we briefly describe some of the additional error mitigation methods used as part of our experimental results. On top of the zero-noise extrapolation methods described in the main text, we also employ Pauli twirling \cite{WallmanRandomizedCompiling, Ware_2021, Kim2023}, dynamical decoupling \cite{Ezzell2023} and twirled readout error extinction (TREX) \cite{PhysRevA.105.032620} as error mitigation methods to improve our circuits' resilience to hardware noise.

Pauli twirling consists of randomly generating a set of four single qubit Pauli gates around each two-qubit gate such that its logical expression remains unchanged up to a global phase. Randomly sampling all possible conjugations across all two-qubit gates present in the circuit will diagonalize their associated noise channels in the Pauli basis of the super-operator representation and has been experimentally shown to decrease the worst-case error rate and make the noise easier to model. In practice, for our experimental results we generate 32 random twirled circuits per noise-scaled ZNE circuit and average over their measured expectation values. 

Dynamical decoupling (DD) on the other hand does not require generating more circuits. Instead, for every circuit which is run, whenever a qubit is left idle a sequence of single qubit gates are applied followed by their inverses, with some schedule of delays in between which fills the idle time. This protects against environmentally induced dephasing and decoherence, and has been found to reduce cross-talk between gates \cite{doi:https://doi.org/10.1002/9781118742631.ch11, PhysRevApplied.20.064027, Kim2023QEM}. 
We use a minimal sequence of two ${\text{Pauli-}X}$ gates for DD in our circuits.

In order to mitigate readout errors we utilized Twirled Readout Error eXtinction (TREX) which effectively diagonalizes the readout error map through random application of Pauli strings in $\{I,X\}^{\otimes N}$ to the qubits, followed by inversion in post processing, at the cost of running one set of calibration circuits. We do not run benchmark circuits for these calibration circuits, as they contain only one layer of single qubit gates.

The classical extrapolation for ZNE and bnZNE is performed with respect to both linear and exponential curve fits. Typically, we try both fits and take the best-performing option with respect to the benchmark circuits, thus giving allowing us to choose a fitting independently of any a-priori knowledge of the true, and in utility-scale calculations, unknown value.

\section{Software}
\label{app:software}

Alongside this work, we provide software \cite{Github} capable of performing our newly developed error mitigation techniques for arbitrary applications. We also include the raw hardware data produced over the course of this work.

The software supports all of the error mitigation techniques used in this work, namely: zero-noise extrapolation (ZNE), benchmarked-noise zero-noise extrapolation (bnZNE), inverse-circuit ZNE (IC-ZNE) Pauli twirling (PT), dynamical decoupling (DD) and twirled readout error extinction (TREX). There are several customisation options available which we outline in Table \ref{table:qem}. We also include two tutorial notebooks outlining how to generate circuits for the IBM hardware and how to post-process the measurement results.

\begin{figure*}[]
    \centering
    \includegraphics{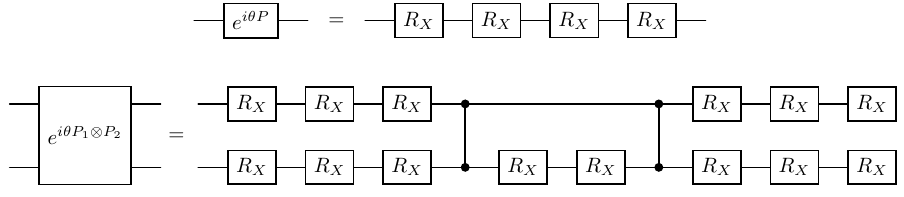}
    \caption{Example of our consistent transpilation method for Pauli rotations applied to the IBM superconducting hardware with basis gate set $\{\text{CZ}, R_Z(\theta), X, \sqrt{X}\}$. We find any single-qubit Pauli rotation can be transpiled using four $R_X$ ($X$ or $\sqrt{X}$) gates, and any two-qubit Pauli rotation using fourteen $R_X$ gates and two $\text{CZ}$ gates, plus arbitrarily many error-free $R_Z$ gates in either case.}
    \label{fig:rigid-compilation}
\end{figure*}

\section{Consistent transpilation of Pauli rotations for hardware-agnostic benchmark circuits on IBM hardware}\label{app A}

We briefly discuss our implementation of the hardware-agnostic benchmark circuits on the IBM hardware with basis gate set $\{\text{CZ}, R_Z(\theta), X, \sqrt{X}\}$, which was used for the circuits in Figure \ref{fig:numerical-results}. This amounts to producing a consistent method of transpiling single and two-qubit Pauli rotations such that any rotation contains the same numbers of erroneous single and two-qubit native gates acting on the same qubits.

Recall that the $\sqrt{X}$ and $X$ gates are implemented via equal-time pulses and are reported to have equal error rates \cite{ibm_quantum}, whilst the $R_z$ gate is simulated virtually and can be viewed as error-free \cite{PhysRevA.96.022330}. Hence, we ensure that each Pauli rotation contains the same number of erroneous $R_X$ ($X$ or $\sqrt{X}$) gates in the same places, interleaved with arbitrary numbers of $R_Z(\theta)$ gates. 

We illustrate our method of doing this in Figure \ref{fig:rigid-compilation}. We find for example that any single-qubit Pauli rotation $e^{i\theta P}$ may be transpiled using four $R_X$ gates and an arbitrary number of $R_Z$ gates. Similarly, any two-qubit Pauli rotation $e^{i\theta P_1 \otimes P_2}$ may be transpiled using fourteen $R_X$ gates. In practice, this is done by compiling the gate using the Qiskit transpiler \cite{qiskit2024} and inserting any additional $R_X$ gates as required. Hence, this method of transpilation provides a consistent strategy for producing benchmark circuits with the same gate layout, given any application circuit described in terms of single and two-qubit Pauli rotations. This approach may then be repeated for any given choice of hardware, provided consistent transpilation is possible with the native basis gate set according to the conditions set out in section \ref{sec:hardware-agnostic}. 
\begin{figure*}[]
    \hspace*{-1em}
    \includegraphics[scale=0.9]{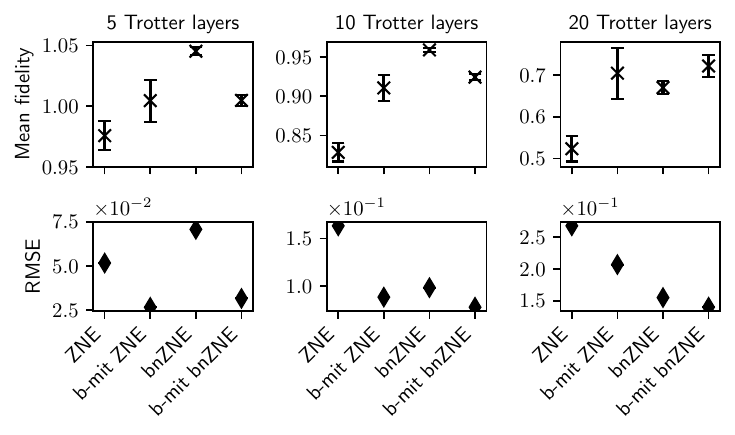}
    \caption{
    Experimental 100-qubit results using \textit{entangling} benchmark circuits (appendix \ref{sec:entangling-benchmarks-for-trotterized-circuits}) with (a) ZNE, (b) bias-mitigated ZNE, (c) bnZNE, and (d) bias-mitigated bnZNE. Compared to the unentangling circuits, we observe greater hardware noise in these results. However, we also observe greater improvement to the mean fidelity and reduction in RMSE between ZNE and the best-performing method compared to the results in the main text.
    }
    \label{fig:entangling-bms}
\end{figure*}

\begin{figure*}[]
    \hspace*{-1em}
    \includegraphics[scale=0.9]{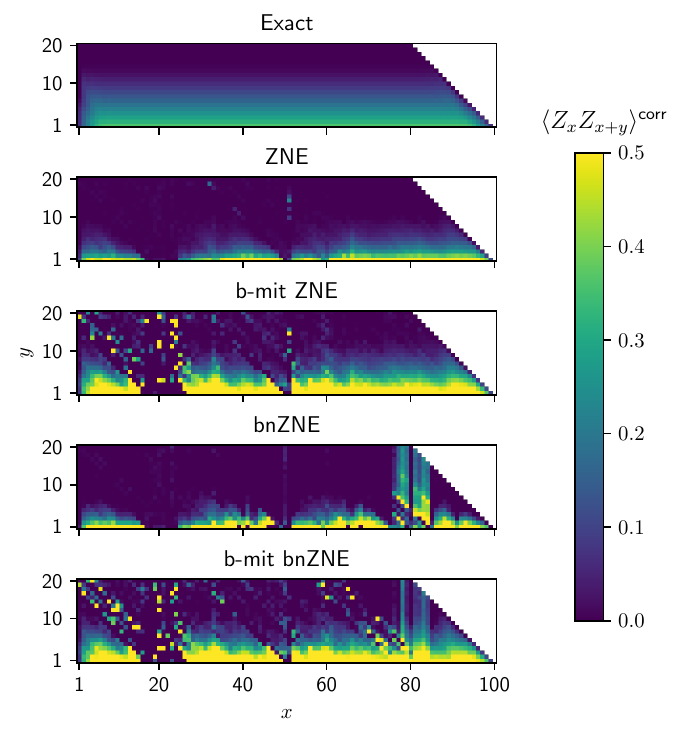}
    \caption{
    Experimental measurement of the two-qubit correlations $\braket{Z_xZ_{x+y}}^\text{corr}$ for a 100-site one-dimensional kicked Ising chain ran on the \texttt{ibm\textunderscore fez} superconducting device using entangling benchmark circuits. We find in general that bias-mitigation helps to recover the long-range correlations. We also observe the effects of problematic qubits in the chain. 
    }
    \label{fig:entangling-bms-correlations}
\end{figure*}

\section{Entangling benchmark circuits for repeated-layer applications}
\label{sec:entangling-benchmarks-for-trotterized-circuits}

Here, we briefly describe an additional method of benchmark circuit generation which ensures the benchmark circuits are at least partially entangling, with the motivation of capturing more of the device noise behaviour compared to the standard benchmark circuits which remain a product state throughout. We also present the corresponding experimental hardware results. We assume as input an application circuit obeying the following properties:
\begin{itemize}
    \item The application consists of $2L$ layers where each layer contains a consistent layout of erroneous single and two-qubit gates.
    \item Each layer can be inverted to produce a new layer containing the same consistent layout of erroneous single and two-qubit gates.
\end{itemize}
This method is specifically targeted to Trotterized circuits where each layer may correspond to a series of Pauli rotations which can be straightforwardly inverted. Technically, however, one could assume $L=1$ without loss of generality. 

Our method of generating benchmark circuits is as follows:
Given an application circuit 
\begin{equation}
    \mathcal{A} = U_{2L} U_{2L-1} \cdots U_1
\end{equation}
where each of the $2L$ layers $U_i$ obeys the properties above, we define the corresponding benchmark circuit to be
\begin{equation}
    \mathcal{B} = \left( U_1^{-1} U_2^{-1} \cdots U_L^{-1} \right) \left( U_L U_{L-1} \cdots U_1 \right).
\end{equation}
In other words, we apply the first $L$ layers of the application circuit followed by their inverses to produce a circuit which is logically equivalent to identity. Since Trotterised circuits typically consist of layers of transpiled Pauli rotations, it should be straightforward to ensure consistency between the gate layouts of the two transpiled circuits.

In Figures \ref{fig:entangling-bms} and \ref{fig:entangling-bms-correlations} we present experimental measurements of the single-site magnetisations $\braket{Z_i}$ and two-qubit correlations $\braket{Z_x Z_{x+y}}^\text{corr}$ as done in the main text. Interestingly, we observed a significantly degraded device behaviour for this experiment compared to the results in the main text. One likely contributor to this issue is the presence of problematic (highly erroneous) qubits on the hardware, which can be identified from Figure \ref{fig:entangling-bms-correlations} as those with positions close to qubits 20 and 50 in the chain. We analyse this further in appendix \ref{sec:identification-of-hardware-imperfections}. Nonetheless, we observe significant improvements versus standard ZNE in the measurement of these expectation values. In particular, the reduction in RMSE between ZNE and the best-performing method is greater for this case when compared to the unentangling benchmark circuits in Figure \ref{fig:numerical-results-hardware}: here we observe reductions in the RMSE of approximately 0.025, 0.07, 0.11 for 5, 10, 20 Trotter layers respectively compared to 0.005, 0.02, 0.05 before. This indicates that these entangling benchmark circuits may perform better than the hardware-tailored unentangling benchmark circuits for bias-mitigation under equal noise conditions. 

We also note here the work of Kim et al. \cite{kim2025enhancedextrapolationbasedquantumerror}, in which ZNE for layered-structure circuits is performed by benchmarking the fidelity of a single layer by running the layer followed by its inverse. In terms of scalability, this work offers an advantage over our approach in that the full circuit depth is not required for the benchmarking case but rather only twice the depth of a single layer. However, like inverted-circuit ZNE \cite{PhysRevA.110.042625}, this method requires measurement of the all-zero state, the probability of which decays exponentially with the number of qubits (see also the next section). Additionally, as the authors state, their method is limited to calculating observation probabilities rather than generic expectation values. Nonetheless, combined instead with our approach for scalable estimation of the error rate in bnZNE (equation (\ref{eq:bnzne-error-rate})), this method could perform very well for deep applications with many similar layers. 

\section{Inverted-circuit zero-noise extrapolation (IC-ZNE)}
\label{sec:ic-zne}

\begin{figure*}[t]
    \hspace*{-1em}
    \includegraphics[scale=0.9]{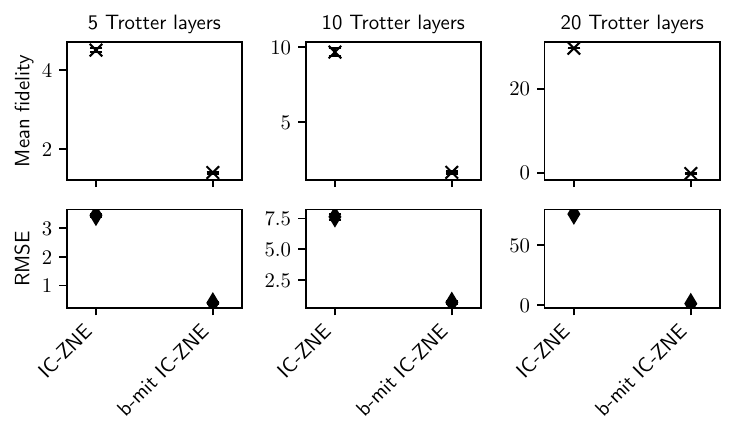}
    \caption{
    Experimental 100-qubit results using (a) IC-ZNE and (b) bias-mitigated IC-ZNE. We find IC-ZNE does not produce accurate expectation values, but that our bias-mitigation approach is able to substantially improve them.
    }
    \label{fig:icZNE}
\end{figure*}
\begin{figure*}[]
    \hspace*{-1em}
    \includegraphics[scale=0.9]{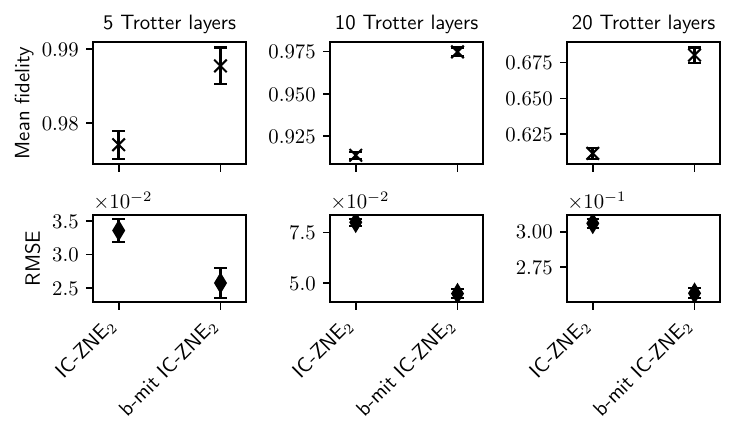}
    \caption{
    Experimental 100-qubit results using (a) IC-ZNE$_2$ and (b) bias-mitigated IC-ZNE$_2$, where IC-ZNE$_2$ denotes standard IC-ZNE combined with our modified scalable error calculation method. Compared to IC-ZNE (Figure \ref{fig:icZNE}), we see a clear improvement in the quality of results with realistic fidelities close to one. At 20 Trotter layers we observe the limited scalability of this method with circuit depth, since the IC-ZNE circuits require a doubled circuit depth and thus noise saturates much faster than for our bnZNE method. 
    }
    \label{fig:icZNE-modified-errors}
\end{figure*}

In this section, we discuss one of the motivations for our bnZNE method, namely inverted-circuit ZNE (IC-ZNE) \cite{PhysRevA.110.042625} and how our work may be viewed as addressing its scalability limitations.

The IC-ZNE method aims to improve calculation of the error rate at each standard ZNE level. Instead of extrapolating the datapoints $(r, \braket{O}(r))$ to $r=0$ as in standard ZNE, we instead extrapolate datapoints  $(\varepsilon(r), \braket{O}(r))$, to $\varepsilon=0$, where $\varepsilon(r)$ is the inferred noise level at ZNE level $r$. Unlike our bnZNE method, this value is obtained by running the noise-scaled application circuit followed by its inverse and measuring the probability $p_0(r)$ of obtaining the initial $\ket{0}^{\otimes n}$ state, where $n$ is the number of qubits. $\varepsilon(r)$ is then a function of $p_0(r)$. We present the results of running IC-ZNE for our 100-qubit kicked Ising circuits on the IBM hardware in Figure \ref{fig:icZNE}. As we will now discuss, we found that IC-ZNE led to very inaccurate results even at 5 Trotter layers, with the mean fidelity more than four times the true value. Nevertheless, we found our bias-mitigation approach was able to make a substantial improvement in results, even at 20 Trotter layers.

Ultimately, we believe that IC-ZNE suffers from two central limitations to its scalability. Firstly, determining the error strength requires measuring the probability of observing the all-zero state. This quantity decays exponentially with both circuit depth and the number of qubits such that for utility-scale circuits it is impractical to measure. To that end, we can improve the scalability by instead applying our scalable error calculation method as used for bnZNE. In other words, for some target observable $O$ acting non-trivially on a subset of qubits $\mathcal{Q}$, we can define
\begin{equation}
    P_0 \equiv \prod_{q \in \mathcal{Q}} (1-p_0(q)),
\end{equation}
where $p_0(q)$ is the measured probability of measuring the single-qubit zero state on qubit $q$. Instead of exponentially decaying to zero with the number of qubits, these probabilities will ultimately saturate to $\frac{1}{2}$ with the noise level and thus with circuit depth. This means that $P_0$ can be measured more precisely without requiring exponentially many shots. We can then apply the usual calculation of the error rate $\varepsilon$ from $P_0$ used in the original work, namely \begin{equation}
    \varepsilon = \begin{cases}
    \frac{1-\sqrt{P_0 - \frac{1-P_0}{2^n}}}{1+\frac{1}{2^n}} & P_0 > \frac{1}{2^n}; \\
    \frac{1-P_0}{1+P_0} & P_0 \leq \frac{1}{2^n}
    \end{cases}
\end{equation}
where $n$ denotes the number of qubits in the circuit. We applied this procedure to the same 100-qubit experiment as above and present the results in Figure \ref{fig:icZNE-modified-errors}, where we denote the procedure as IC-ZNE$_2$. Compared to standard IC-ZNE, we see a clear improvement in the quality of results, obtaining expectation value fidelities close to unity for 5 and 10 Trotter layers. We also observe that our bias mitigation procedure improves the fidelity and RMSE in each case. Notably, at 5 and 10 Trotter layers the bias-mitigated values slightly outperform the best obtained values from our bnZNE method in Figure \ref{fig:numerical-results-hardware}. This is to be expected, since the error-detection circuits in the IC-ZNE case are entangling, using precisely the application circuit and its inverse, whilst our benchmark circuits are structure-mirroring but unentangling and functionally different.  

At 20 Trotter layers we appear to observe the second scalability limitation of IC-ZNE. Since the IC-ZNE error detection circuits require twice the circuit depth of the application circuit, we expect the device noise to saturate more quickly with application depth and the quality of results to be impacted. In this case this leads to expectation value fidelities around 0.65. Our bnZNE method fixes this limitation by using error-detection circuits which have the same depth as the application, leading to an equivalent fidelity above 0.9 in Figure \ref{fig:numerical-results-hardware}.

\begin{figure*}[t!]
    \hspace*{-1em}
    \includegraphics[scale=0.9]{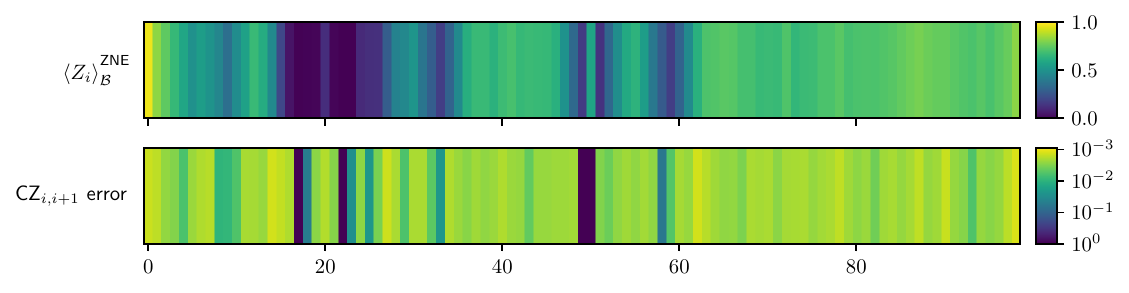}
    \caption{Comparison between the single-site expectation values $\braket{Z_i}_\mathcal{B}^\text{ZNE}$ measured under standard ZNE with the native two-qubit gate error rates reported by the backend. We observe a clear correspondence between the two.}
    \label{fig:error-detection}
\end{figure*}

\section{Identification of hardware imperfections}
\label{sec:identification-of-hardware-imperfections}

A final feature of our bias-mitigation procedure that we observed is the ability to better identify hardware imperfections using the benchmark circuits. The most clear example of this is in Figure \ref{fig:entangling-bms-correlations}, where the bias-mitigation recovers the correlations for long-range interactions and clearly shows a breakdown in the quality of results within the lightcones of qubits with indices close to 20 and 50. This is analagous to previous work which has argued that deviations in structured correlation functions across phase diagrams could be used to assess trustworthy qubits in a given experimental result for condensed matter models \cite{Lively2024}. More generally, we find that the measured single-site expectation values of the benchmark circuits can be used to actively identify problematic qubits within the hardware layout. In Figure \ref{fig:error-detection}, we compare the single-site expectation values of the benchmark circuit $\braket{Z_i}_\mathcal{B}^\text{ZNE}$ measured under standard ZNE with the native two-qubit gate error rates reported by the backend at the time of execution, observing a clear correspondence between the two. Our benchmark circuits could hence be used to automatically detect issues with the corresponding device, particularly in the case where recently benchmarked error rates are not available and full benchmarking would be too costly.

\end{document}